\documentclass[12pt,preprint]{aastex}

\newcommand{\myemail}{rmason@ctio.noao.edu}
\def\deg{\hbox{$^{\:\circ}$~}}
\def\mic{\hbox{$\mu$m}}

\shorttitle{Mid-IR spectroscopy of NGC~1068}

\shortauthors{Mason et al.}

\begin{document}
\title{ Spatially-resolved mid-infrared spectroscopy of NGC~1068: the nature and distribution of the nuclear material}

\author{R. E. Mason}
\affil{NOAO Gemini Science Center, Cerro Tololo Interamerican Observatory, Casilla 603, La Serena, Chile}
\email{\myemail}

\author{T. R. Geballe}
\affil{Gemini Observatory, 670 N. Aohoku Place, Hilo, HI 96720, USA}
\email{tgeballe@gemini.edu}

\author{C. Packham}
\affil{Department of Astronomy, University of Florida, PO Box 112055, 211 Bryant Space Center, Gainesville, Fl 32611 USA}
\email{packham@astro.ufl.edu}

\author{N. A. Levenson}
\affil{Department of Physics and Astronomy, University of Kentucky, 177 Chemistry/Physics Building, Lexington, KY 40506, USA}
\email{levenson@pa.uky.edu}

\author{M. Elitzur}
\affil{Department of Physics and Astronomy, University of Kentucky, 177 Chemistry/Physics Building, Lexington, KY 40506, USA}
\email{moshe@pa.uky.edu}

\author{R. S. Fisher}
\affil{Gemini Observatory, 670 N. Aohoku Place, Hilo, HI 96720, USA}
\email{sfisher@gemini.edu}

\and

\author{E. Perlman}
\affil{Joint Center for Astrophysics, Physics Dept., University of Maryland, 1000 Hilltop Circle, Baltimore, MD 21250}
\email{perlman@jca.umbc.edu}

\begin{abstract}

We present spatially-resolved, near-diffraction-limited  10~$\mu$m spectra of the nucleus of the Seyfert 2
galaxy NGC~1068, obtained with Michelle, the mid-IR imager and spectrometer on
the 8.1 m Gemini North telescope.  The spectra cover the nucleus and the central
6.0\arcsec\ $\times$ 0.4\arcsec\ of the ionization cones at a spatial
resolution of approximately 0.4\arcsec\ ($\approx$30 parsecs). The spectra extracted in 0.4\arcsec\ steps
along the slit reveal striking variations in continuum slope, silicate
feature profile and depth, and fine structure line fluxes on subarcsecond scales,
illustrating in unprecedented detail the complexity of the circumnuclear
regions of this galaxy at mid-IR wavelengths. A comparison of photometry in various
apertures reveals two distinct components: a compact (radius $<$ 15 pc), bright
source within the central 0.4\arcsec\ $\times$ 0.4\arcsec\ and extended,
lower brightness emission. We identify the compact source with the AGN
obscuring torus, and the diffuse component with the AGN-heated dust in the
ionization cones. While the torus emission dominates the flux observed in the
near-IR, the mid-IR flux measured with apertures larger than about 1\arcsec\ is
dominated instead by the dust emission from the ionization cones; in spite of
its higher brightness, the torus contributes less than 30\% of the 11.6~\mic\
flux contained in the central 1.2\arcsec\ region. Many previous
attempts to determine the torus spectral energy distribution are thus likely to be significantly affected by contamination from the extended emission. The observed spectrum of the compact source is compared with clumpy
torus models, the first detailed comparison of such models with observational
data. The models require most of the mid-IR emitting clouds to be located
within a few parsecs of the central engine, in good agreement with recent
mid-IR interferometric observations. We also present a UKIRT/CGS4 5 $\mu$m
spectrum covering the R(0) -- R(4) lines of the fundamental vibration-rotation
band of $^{12}$CO. None of these lines was detected, and we discuss these
non-detections in terms of the filling factor and composition of the nuclear
clouds.

\end{abstract}


\section{Introduction}
\label{intro}

The advent of the unified model of active galactic nuclei \citep[AGN;
e.g.][]{Antonucci93}, with its torus of dusty material,  helped make sense of many
of the observed properties of the numerous classes of active galaxy, but
has brought with it its own set of challenges. An intense modeling
effort has been under way with the aim of understanding and predicting the emission from
dusty tori and explaining the observational data
\citep{PK92,PK93,Granato94,EHY95,Manske98,Nenkova,vanBem,Schartmann05},
and with some success in matching the gross infrared properties of AGN,
but several observations have proven particularly difficult to reproduce.
Especially problematic have been the width of the IR spectral energy
distributions (SEDs), the similarity of the observed SEDs given the wide
range of variation in X-ray column densities, and, most notoriously, the
behavior of the 9.7 $\mu$m silicate feature.  While the feature is
commonly seen in absorption in type 2 objects \citep{Roche91,Sieb04}, the
corresponding emission feature expected for type 1 objects has to date
only been detected in a handful of objects, mostly QSOs
\citep{Hao05,Siebenmorgen05,Sturm05}. Torus models must therefore be
capable of producing a silicate feature in type 2 AGN while suppressing it
in most type 1 objects.

Recently, progress has been made towards overcoming some of these
problems. For instance, using a model in which the dust is contained in
discrete clouds, thought to be necessary to ensure dust survival in the
harsh AGN environment, \cite{Nenkova} were able to construct SEDs broad
enough to match typical observed SEDs. In these models, the silicate
absorption feature was visible in an edge-on view of the torus but the
emission could be greatly weakened in a face-on view. By biasing the grain
size distribution to large sizes, \citet{vanBem} were able to suppress the
silicate emission feature, although their models tend to predict larger
torus inner radii ($\ge$10 pc) than recent observational limits (see
below). \citet{Schartmann05} were also able to suppress the silicate
emission and fit the mean type 1 SED by introducing different sublimation
radii for large and small grains. Given advances such as these, torus
models will benefit from high-quality observations against which they can
be thoroughly tested.

Recent high spatial resolution observations of several active galaxies at
10 and 20 $\mu$m, where a torus re-emits a substantial fraction of the
radiation incident upon it, indicate that the tori may be very small. For
example, the outer diameter of the mid-IR emitting dust associated with the
torus has been shown to be no greater than a few parsecs in NGC~1068 and
the Circinus galaxy \citep{Jaffe,Packham05}, and $\le$35 pc in NGC~4151
\citep{Radomski03}, although cooler dust at larger distances is not ruled out. At the same time, some AGN are also associated with
extended mid-IR emission from dust on scales of a few hundred parsecs or
larger \citep{Maiolino95, Krabbe01, Radomski02, Gorjian04, Sieb04}. On
moving from large to small apertures, the IR spectrum of an active galaxy
will often change from being dominated by PAH emission bands to showing a
featureless continuum or silicate absorption feature
\citep{LeFloch,Soifer02,Sieb04}. Where the AGN component can be spectrally isolated
from large-aperture data, the expected clear dichotomy in hard
X-ray/mid-IR flux ratios between types 1 and 2 AGN is not observed,
perhaps because dust heated by the nuclear source but not associated with
the torus itself is contributing to the mid-IR flux
\citep{Lutz04}. It is
therefore clear that observations at the highest possible spatial
resolution will be necessary to disentangle the torus and extended
emission and provide a fair test for models of the dusty torus.

As one of the nearest, brightest and best-studied AGN, the archetypal
Seyfert 2 galaxy, NGC~1068 (14.4 Mpc distant for $\rm H_{0}=75~km \ s^{-1}
\ Mpc^{-1}$; 1\arcsec\ = 72 pc), is an excellent candidate for mid-IR observations at high
spatial resolution. The existence of a hidden broad line region in
NGC~1068, as revealed by the broad emission lines detected in polarized
light \citep{AM85}, along with evidence for high extinction along our line of sight, strongly suggests the presence of obscuring dust in a
toroidal geometry. Further evidence for a torus comes from the strong
nuclear mid-IR source coupled with emission from ionized gas extended in a
roughly conical structure oriented approximately northeast - southwest
\citep[e.g.][]{Macchetto94}. Modeling of 10~$\mu$m interferometric
observations of the nucleus suggests a hot, compact ($\le$1 pc) structure,
surrounded by a more extended (approx. 2$\times$3 pc) 
warm component, interpreted as dust in the hot inner wall of the torus
\citep{Jaffe}. Larger-scale 10 and 20 $\mu$m emission from NGC~1068 is
observed to extend out from the central source to the northeast and
southwest \citep{Braatz93,Cameron93,Alloin,Galliano05}; further out, low
surface brightness mid-IR emission is detected from a ring of star
formation located at kiloparsec distances from the
nucleus \citep{Telesco88}.

The 9.7 $\mu$m silicate feature is seen in absorption in NGC~1068
\citep{Roche84,Jaffe,Sieb04,Rhee05}. Previous narrow-band imaging suggests that
the depth of the absorption varies over the nuclear region
\citep{Bock2,Galliano}, but the spatial coverage and spectral resolution
of those studies has been insufficient to allow detailed conclusions about
the distribution and profile of the feature to be drawn.  We have
therefore performed spatially resolved N band spectroscopy of the
nucleus and ionization cones of NGC~1068 with Michelle, the mid-IR imager
and spectrometer on the 8.1 m Gemini North telescope. The spatial resolution of
the spectra, approximately 0.4\arcsec, allows us to separate the nuclear
and extended emission regions and reveals the distribution of the silicate feature
across the nuclear region.  Having largely isolated the spectrum of the
torus in the spectrum of the central 0.4$\times$0.4\arcsec\ ($\approx$
30$\times$30 pc) of NGC~1068, we compare it with the clumpy torus model of
\citet{Nenkova}. The data set also allows us to examine the detailed
spatial distribution of fine structure emission lines in the 10~$\mu$m
region, which have been reported previously only in very large aperture
ISO measurements \citep{Lutz00} or in few-arcsecond aperture ground-based
spectroscopy \citep{Sieb04}.

In \S\ref{obs} we describe the observations and reduction of the data,
while \S\ref{results} presents the results obtained for the acquisition
images, nuclear spectra and model comparison, extended silicate feature and
emission lines. We also report the results of a sensitive search at
5~$\mu$m for CO lines towards the nucleus of NGC~1068, which serves as a
probe of the nature and distribution of molecular gas near the center
of this galaxy. The implications of this work are discussed in
\S\ref{discuss} and summarized in \S\ref{conc}.

\section{Observations and data reduction}
\label{obs}

\subsection{N-band spectroscopy}

N-band acquisition images and spectra of the nucleus of NGC~1068 were
obtained during commissioning of Michelle \citep{Glasse97} on Gemini North.  Observations were performed on UT
2004 August 10 (epoch 1) and repeated on 2005 January 31 (epoch 2).  In
each case the 2-pixel (0.4\arcsec) wide slit, oriented at 20\deg east of north,
was used together with the lowN grating, resulting in a dispersion of
0.024~$\mu$m/pixel and R$\sim$200.  Observations were made using a standard
chop-nod technique to remove time-variable sky background, telescope
thermal emission and to reduce the effect of 1/f noise from the array/electronics.  The chop throw was
15$\arcsec$, fixed at 20$\degr$ east of north, and the target was nodded the same
amount along the slit.  Extended emission has been detected at $\sim$15\arcsec\ from the nucleus, but its surface brightness \citep[usually $<<$10 mJy/arcsec$^{2}$;][]{Telesco88} is negligible compared to that of the region from which spectra were extracted. As fast tip/tilt guiding is possible in only the on-source beam, the data used were on-source only and the quoted exposure
times are for the on-source data.

In both epochs we obtained acquisition images with on-source
integration times of 4 seconds through the Si-5 11.6$\mu$m filter
($\Delta$$\lambda$ = 1.1$\mu$m, 50$\%$ cut-on/off).  In spectroscopic mode,
on-source integration times were 1043 s for the first epoch and 680 s for
the second.  The plate scale of Michelle is 0.099\arcsec/pixel in imaging
mode and 0.198\arcsec/pixel in spectroscopic mode.  The weather was good
for both epochs and the FWHM of the standard star was measured
to be 0.35\arcsec\ on the first and 0.43\arcsec\ on the second. The
diffraction limit (1.22$\lambda$/D) of the telescope at the cut-on of the Si-5 filter is 0.34\arcsec.

The orientation of the slit on the galaxy, and the area from which spectra
were extracted, are shown in Fig.~\ref{fig:slitpos}. The figure illustrates
that the slit was aligned to the peak brightness of NGC~1068 to better than
0.1\arcsec\ on the second epoch and to around 0.2\arcsec\ on the first, with the difference attributable to slight pointing errors during instrument commissioning. To
investigate the possibility that the object drifted out of the
spectroscopic slit during the observation, due to effects such as guiding
errors or differential refraction between the optical guide camera and the
mid-IR science beam, we split the observation into several sections and
measured the peak emission in each.  The counts changed by $\le$6\% and
$\le$10\% on the first and second nights respectively, typical of MIR
photometric variations \citep[e.g.][]{Radomski02,Packham05}, implying that
any drifts in guiding were less than 0.1\arcsec.

Michelle data files contain planes consisting of the difference for each
chopped pair for each nod-set.  Using the Gemini IRAF package, these
difference images were combined until a single frame was created.  During
this process the chopped pairs were examined to identify any obviously
compromised by anomalously high background, electronic noise, or other
problems; it was not judged necessary to exclude any frames from the final
images.  Finally, the resultant image was divided by a spectroscopic flat
field.  The spectra were extracted in 0.4\arcsec\ sections (2 pixels along
the slit in spectroscopic mode, corresponding well to the average seeing
disc over the two epochs), covering the central few arcseconds of
NGC~1068, stepping outwards from an aperture centered on the flux peak.
Michelle spectra show a slight ($<$ 1 pixel) curvature across the array,
which was measured using the standard star data and the resulting
apertures used in extracting the NGC~1068 spectra. The spectra were
wavelength calibrated to an accuracy of 0.02~$\mu$m using the telluric
absorption lines present in the peak NGC~1068 spectrum, and then divided
by the standard star (BS 1101, F9V) and multiplied by a 6000 K blackbody
curve.

Cancellation of the deep telluric O$_{3}$ band in the spectra extracted at
distances $\ge$4 pixels (0.8\arcsec) from the nucleus was poor, so a
spectrum of the O$_{3}$ feature (derived from the standard star spectra
set to zero outside the band) was added back into the star spectra in
varying amounts and these ``composite'' star spectra used to divide out
the telluric band in those NGC~1068 spectra. The scaling of the O$_{3}$
band in the star spectra was determined by comparing the depth of the
resulting, ratioed NGC~1068 spectra at 9.7~$\mu$m with that of a spline
fit to the ratioed NGC~1068 spectra, with the 9.3-10.0~$\mu$m region
excluded from the fit.

Flux calibration was achieved through comparison with BS~1101, whose N
band magnitude was taken to be
2.9\footnote[1]{www.gemini.edu/sciops/instruments/michelle/MichSpecStd.html}. The resulting spectral flux densities are expected to be accurate to about $\pm$20\%.
The calibration agreed to within 40\% between the two epochs, with the
second epoch's data being higher in flux and thus closer to other
small-aperture measurements in the literature \citep{Tomo,Jaffe}.  The
difference in flux between the two epochs is probably mainly due to
the small difference in the positioning of the slit.

Fine structure line emission by the [Ar~III]+[M~VII] blend at 8.99~$\mu$m,
[S IV] at 10.51~$\mu$m, and [Ne II] at 12.81~$\mu$m (rest wavelengths) was readily apparent
out to several arcseconds from the nucleus. Line fluxes were computed from
the averages of 4-6 data points centered on each of the line wavelengths
and five continuum points on either side of each line. The uncertainty in the derived line fluxes is expected to be small except very near the center, where
line-to-continuum ratios are very low and systematic errors dominate
(particularly in deriving the [Ne~II] flux at the edge of the N band window). Apart
from these lines, no other narrow spectral features are apparent in the
reduced spectra.

Slit losses were estimated from the profile of the star along the slit (which was assumed to be circularly symmetric) to
be $\sim$50\% on both nights. The spatial distributions of all three fine
structure lines appear to be fairly smooth, with no major contributions from
pointlike sources, so to correct for systematic overestimates in fluxes resulting from slit losses in the standard star
spectrum used for flux calibration, measured line fluxes at all slit positions were multiplied by a
factor of 0.5. At positions $\ga$6 pixels from the peak, the continuum flux
also approaches a smooth spatial distribution, whereas at the nucleus it
appears dominated by a pointlike source (although somewhat smoothed by
small tracking errors, differential refraction etc. during the long integration). No correction for slit
losses has been applied to the continuum flux density, but the effect of any
correction would be to increase the degree to which it is concentrated on
the nucleus.

\subsection{M-band spectroscopy}

A high resolution M-band spectrum of NGC~1068 was obtained at UKIRT on UT
1998 December 29. The spectrum was acquired with the facility instrument,
CGS4 \citep{Mountain90}, using its 31~l/mm echelle. The slit width was
0.9\arcsec\ and the velocity resolution was 20 km/s (R=15,000).  The total
exposure time was 18.7 minutes. The spectrum of NGC~1068 was ratioed by the
spectrum of HR 804 (A3V), which was observed both before and after NGC~1068
in order to provide a precise airmass match. The pressure-broadened Pf beta
line in HR 804 produces a broad emission bump in the ratioed spectrum,
centered at 4.654~$\mu$m; this was removed by dividing by a heavily
smoothed (Gaussian FWHM = 300 km/s) version of the ratioed spectrum.

\section{Results}
\label{results}

\subsection{The acquisition images}
\label{acq}

\begin{figure}
\includegraphics[width=16cm]{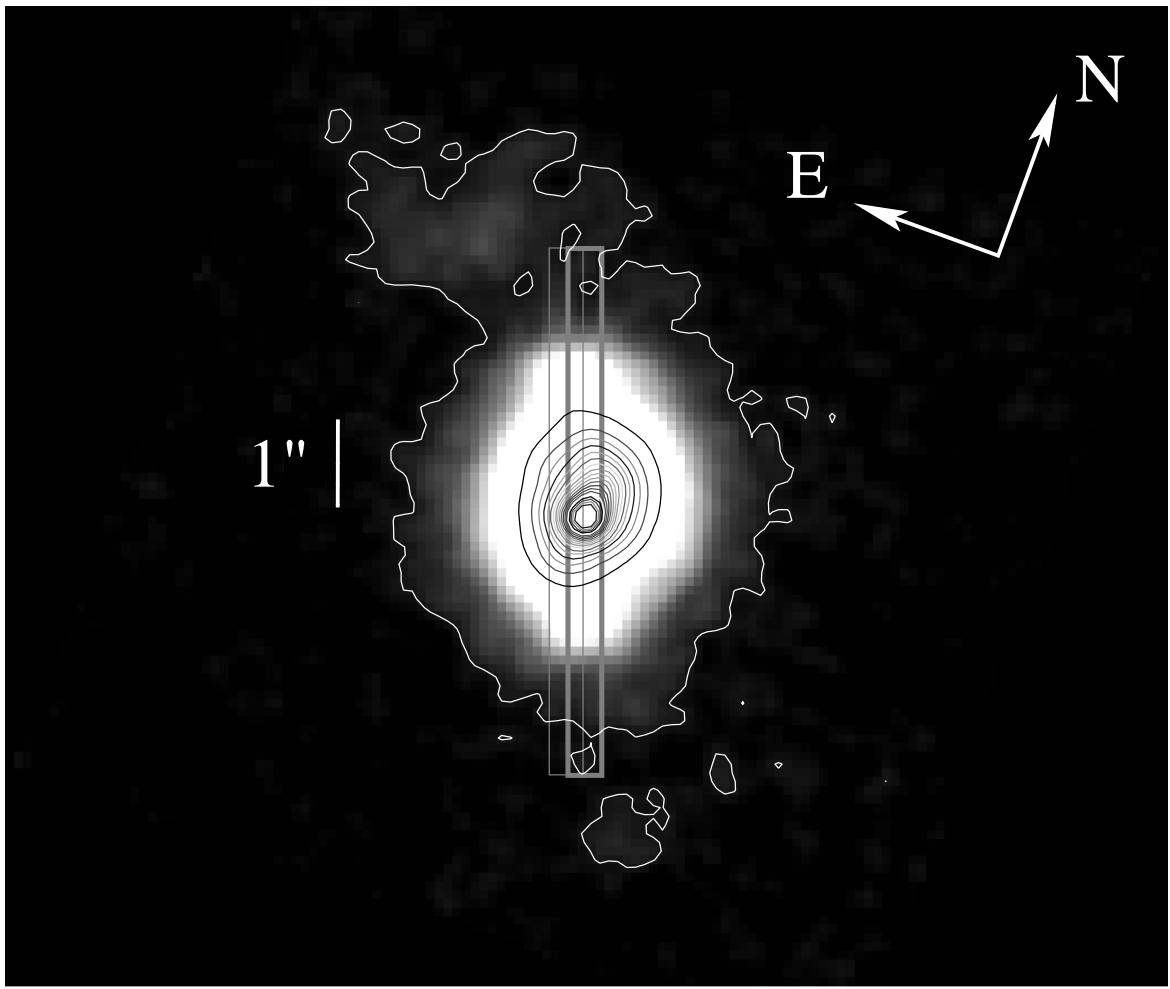}
\caption{Central portion of the 11.6 $\mu$m acquisition image of NGC~1068
(epoch 1) with
the position of the spectroscopic slits for each epoch overlaid (thick
line --- epoch 2; thin line --- epoch 1).  Contours are
linearly spaced at intervals of 50$\sigma$ (starting at 3$\sigma$, or 0.05 Jy/sq. arcsec.).}
\label{fig:slitpos}
\end{figure}

The 11.6~$\mu$m acquisition images (Fig. \ref{fig:slitpos}), taken to
assist in positioning the nucleus in the slit, reveal features similar to
those detected in previous high spatial resolution mid-IR imaging of
NGC~1068.  As shown by \cite{Tomo}, NGC~1068 is initially extended in a
roughly north-south direction centered on the nucleus. The northerly
extension is at a position angle (PA) of $\sim$ 14\deg within $\sim$
0.5\arcsec, turning easterly to a PA of $\sim$20\deg around 1\arcsec\ from
the nucleus. The southern extension is at a PA of 190\deg at 0.5\arcsec\
from the nucleus, bending westerly further from the nucleus. The features
described above agree well with those in the literature, especially those
described by \citet{Bock2}, \citet{Tomo} and \citet{Galliano05}.

Northeast of the nucleus, at around PA=35\deg and 1.8\arcsec\ and
3.7\arcsec\ from the nucleus, extended knots of emission previously
detected by \cite{Alloin} are observed.  However, in our higher spatial
resolution data, we also resolve a low surface brightness
south-southwestern knot of emission just under 4\arcsec\ from the nucleus
at PA=205\degr, also recently detected by \citet{Galliano05}.  The
approximate centers of the northeastern and south-southwestern knots are
located close to a line running radially through the nucleus. The mid-IR
extensions are enclosed within the near-IR polarization structures
observed both north and south of the nucleus by \citet{Packham97}, who
interpreted the structure as scattering from electrons within the
ionization cones.  Clearly the 20\deg position angle of the slit closely
matches the orientation of the regions of extended emission
$\ga$1\arcsec\ from the nucleus described above.

\begin{deluxetable}{ccc}
\tabletypesize{\scriptsize}

\tablecaption{11.6 $\mu$m flux density of NGC~1068 in apertures of various diameters and
the corresponding average surface brightness (flux density/aperture diameter$^2$) }

\tablewidth{0pt}

\tablehead{
\colhead{Aperture diameter} & \colhead{Flux density \tablenotemark{1}} & \colhead{Brightness} \\
\colhead{(arcsec)} & \colhead{(Jy)} & \colhead {(Jy arcsec$^{-2}$)}
}
\startdata
  0.4 &  8.7  &   54 \\
  1.2 &  30   &   21 \\
  1.8 &  39   &   12 \\
  2.0 &  40   &   10 \\
  4.0 &  46   &  2.9 \\
  8.9 &  48   &  0.61\\
\enddata
\label{table1}
\tablenotetext{1}{Flux calibration is expected to be accurate to around 15\%, as is typical at mid-IR wavelengths \citep{Packham05,Radomski03}.}
\end{deluxetable}

In the Gemini images the central core is resolved in the direction along the
ionization cones (PA$\sim$20\degr), with FWHM$\approx$0.5\arcsec, but is
unresolved perpendicular to the cones. This implies  a deconvolved extent of
$\sim$0.3\arcsec\ along the cones and $\le$0.2\arcsec\ perpendicular. That is
similar to the size derived by \citet{Tomo} from deconvolution of Subaru
images. Table~\ref{table1} lists the flux density $F$ measured in various apertures $A$
together with the corresponding beam-averaged surface brightness $F/A^2$. At
the 14.4 Mpc distance to NGC 1068, the observations cover dust emission within
radial distances from $\sim$ 15 pc to $\ga$ 300 pc around the nucleus. The
results 
are in good agreement with the findings
of \citet{Jaffe}, whose reported flux density for a 0.4\arcsec\ aperture is 12 Jy, with
a scatter between 11 Jy and 13.5 Jy. 

Together, the observations reveal the presence of two
distinct emission components: low-brightness emission from dust in the
ionization cones extending to hundreds of pc, and bright emission from a
central compact source, identified as the AGN obscuring torus by \citet{Jaffe}.
The torus is resolved by the interferometric observations, which show that its
mid-IR size is $\sim$ 3pc (0.04\arcsec), much smaller than our angular
resolution. Therefore in our observations the torus is effectively a point
source within the central 0.4\arcsec\ diameter region. However, because the
brightness of this region is much higher than that of the surrounding area, its
spectrum is dominated by the torus flux. Indeed, the mean color temperature
estimated from our flux densities at 8~$\mu$m and 13~$\mu$m in the central
spectrum (\S\ref{MIRspec}) is $\sim$330~K, in good agreement with
\citet{Jaffe}. On the other hand, the flux collected by the larger apertures is
dominated by the ionization cones' lower surface brightness diffuse emission.
For example, in 1.2\arcsec\ and 4\arcsec\ diameter apertures centered on the
nucleus, the core contributes less than 30\%\ and 20\%, respectively, of the
total flux. The extended dust emission from the ionization cones was previously
noted by \citet{Alloin} and \citet{Bock2}, and our observations clarify the
relative importance of the flux from the torus and the cones. Like previous
high spatial resolution mid-IR imaging studies, our images show no evidence for
mid-IR emission extending several tenths of an arcsecond east and west of the
nucleus, as reported by \citet{Marco00} at L and M.



\subsection{The mid-IR spectra}
\label{MIRspec}

\begin{figure}
\epsscale{0.80}
\plotone{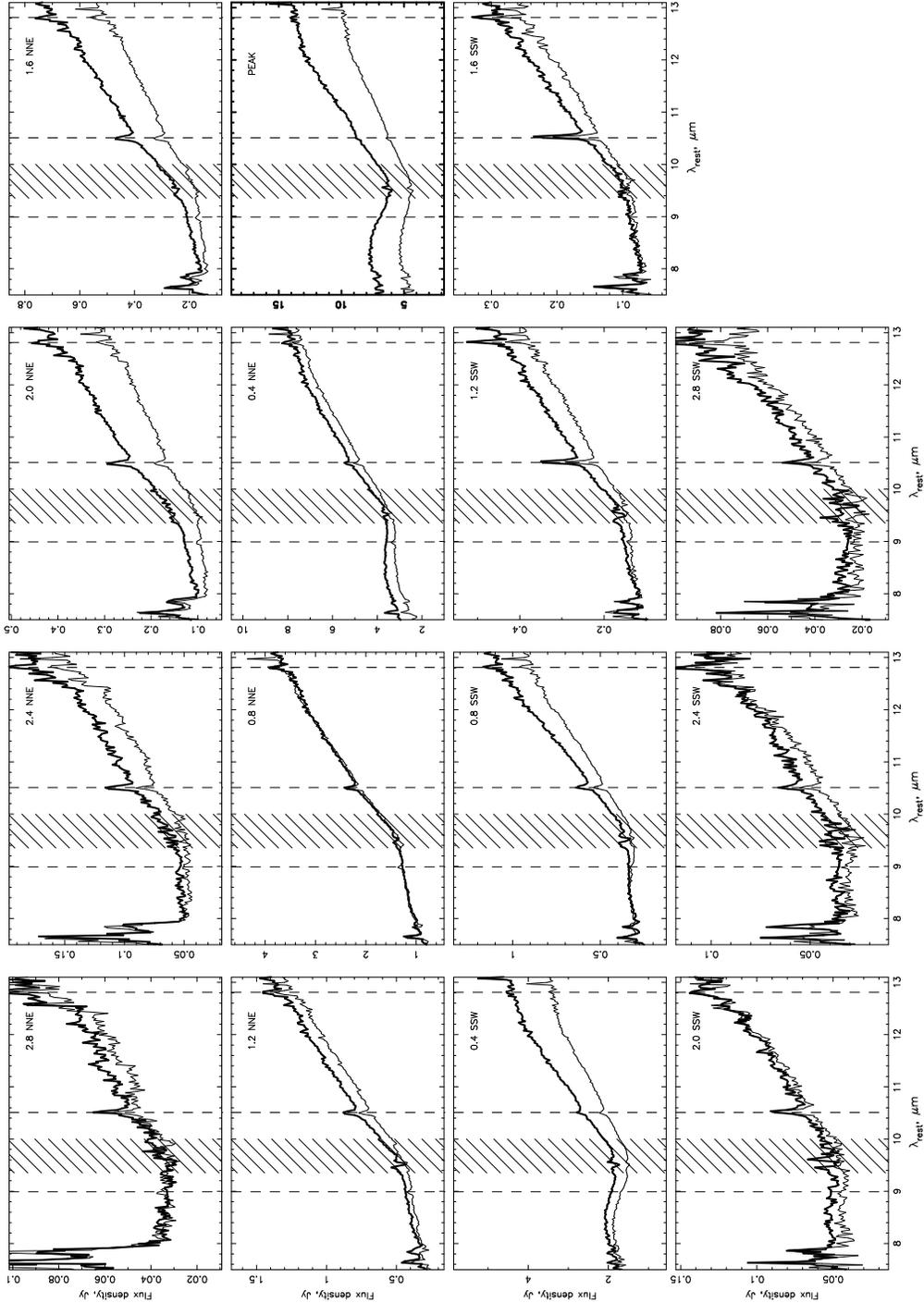}
\caption{Spectra extracted in 0.4\arcsec\ steps over the central
0.4$\times$6\arcsec\ of NGC~1068 along position angle 20\deg (black --- epoch 2; grey --- epoch 1).
Position designations are north-northeast (NNE) and south-southwest
(SSW) of the nucleus with offsets in arcsec of the center of the
0.4\arcsec-wide extraction aperture relative to the mid-IR flux peak. No
correction has been made for slit loss effects, which are different for
the lines and continuum (see \S\ref{obs}).  The dashed lines indicate the positions of the [Ar III/Mg VII] blend at 8.99 $\mu$m and the [S IV] 10.51 $\mu$m and [Ne II] 12.8 $\mu$m fine structure lines, while the hatching shows the approximate extent of the telluric O$_{3}$ band. A bad column partly overlapping the 8.99~$\mu$m emission feature has been interpolated over in the epoch 2 data.} 
\label{fig:all} 
\end{figure}

\begin{figure}[h]
\includegraphics[width=12cm,angle=270]{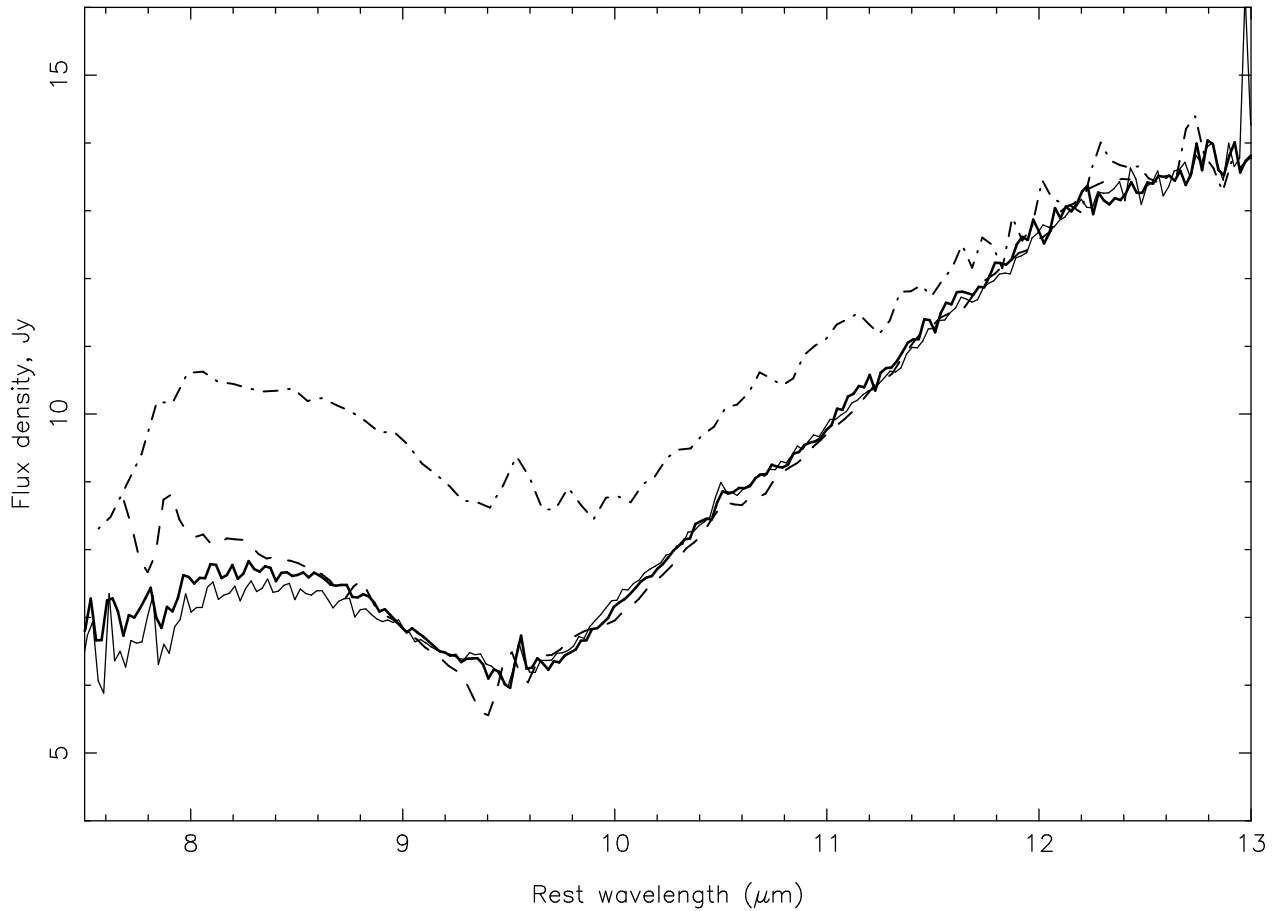}
\caption{Spectra extracted from a 0.4$\times$0.4\arcsec\ region centered on
the flux peak (black --- epoch 2; grey --- epoch 1 $\times$ 1.4). Also shown are the \citet{Rhee05} spectrum of the central 0.25\arcsec $\times$ 0.25\arcsec of NGC~1068 (dashed; scaled to match the Michelle spectrum around 12~$\mu$m) and the single-dish MIDI spectrum of \citet[][dash-dotted]{Jaffe}.}
\label{fig:specs}
\end{figure}

Spectra of the nucleus and ionization cones of NGC~1068, extracted in
0.4\arcsec\ steps (as described in \S\ref{obs}), are shown in Fig.
\ref{fig:all} for both epochs. All of the spectra consist of a strong
dust continuum upon which a broad silicate absorption feature 
and narrow fine structure lines are present. At each slit
position the shapes of the two spectra are similar, showing broadly
similar changes in spectral slope and silicate depth with distance from
the nucleus.  The relative weakness of the epoch 1 spectrum on the
nucleus results from the slit on that date being slightly offset from the
peak of the emission which, unlike the emission from the ionization
cones, is unresolved in the east-west direction. Fig.~\ref{fig:specs}
highlights the good agreement of the profiles of the two central spectra.

The depth of the silicate absorption feature is an important
parameter in modeling of the spectral energy distribution (SED) of AGN
and hence has been studied in detail for NGC~1068 and several other AGN
\citep{Roche91,Sieb04}. However, most previous MIR observations of AGN
have lacked the spatial and/or spectral resolution to probe the nucleus
with minimal contamination from the surrounding MIR-emitting regions.
The observations presented here accomplish this and also provide the
opportunity to spectroscopically study the extended silicate feature
along the ionization cones of NGC~1068, as well as the emission lines
within the cones.

We quantify the depth of the observed silicate feature using $\tau_{9.7}$.
This quantity would be the 9.7~$\mu$m optical depth of an absorbing screen
in front of a background source when the screen emission is negligible. In
realistic configurations the relation of $\tau_{9.7}$ to actual optical
thickness is more complex, but it is still a meaningful phenomenological
indicator. To estimate the 9.7~$\mu$m continuum we interpolated between
the median of the spectrum near 8.2 and 12.4~$\mu$m, wavelengths chosen to
be outside the silicate feature as far as possible without including
regions of poor telluric cancellation or the 12.8~$\mu$m [Ne~II] emission
line. The procedure is equivalent to fitting a linear continuum to the
spectra; the limited wavelength extent of the N-band atmospheric window,
only slightly wider than the silicate feature itself, makes determination
of the true underlying continuum shape difficult, but we believe that this
simple approach is at least indicative.  1$\sigma$ uncertainties were estimated from the scatter of the data points around 8.2, 9.7 and 12.4~$\mu$m used in the calculation of $\tau_{9.7}$.


It has been pointed out by many authors that PAH emission features on the short and long wavelength sides of the silicate absorption feature
(8.6 and 11.3~$\mu$m) can serve to exaggerate the depth of the feature,
particularly when the spectral resolution is very low and/or spatial
resolution is limited. The larger-aperture spectra of \citet{LeFloch} and
\citet{Sieb04} show no evidence of the PAH features, but by themselves
cannot rule out contributions by PAHs in regions as small as the
Gemini/Michelle slit segments. In the high spatial resolution R$\sim$200
Michelle spectra presented here, however, there is no indication, in any
of the spectra, of the PAH features, in particular of the 11.3~$\mu$m
band, which would be easily recognized at this resolving power. The
absence of PAH emission simplifies estimates of $\tau_{9.7}$.  However,
the peak of the silicate band unfortunately coincides with the broad
atmospheric O$_{3}$ absorption band centered at 9.6~$\mu$m, which limits
the accuracy of $\tau_{9.7}$, particularly in spectra far from the nucleus.

In the following sections, we discuss in detail the continuum slope, the
silicate absorption at the nucleus, the more extended silicate feature,
and the emission lines.  We assume throughout that the brightest mid-IR
continuum point marks the buried central engine of NGC~1068.

\subsubsection{Continuum slope}

The slope of the continuum is markedly different within about 0.4\arcsec\ of
the nucleus than at larger distances.  Although there is some variation outside
this region, at all of the outer slit positions the continuum increases
strongly to longer wavelengths. In the central three spectra,
however, the spectral slope in $F_{\nu}$ is significantly less steep, with the bluest continuum
seen at the nucleus. The color temperature of the mid-IR emitting dust, as
estimated from the difference in flux densities at 8~$\mu$m and 13~$\mu$m in
the central spectrum, is $\sim$330~K in the core (as noted above), whereas the
 dust temperature is typically 275~K for distances of 0.8\arcsec\ and
greater. Note that although the radio emission
to the northeast of the nucleus is attributed to synchrotron emission with (in
places) a fairly flat spectrum \citep{Gallimore96b}, the integrated 5 GHz flux density in a 1\arcsec.5 x 1\arcsec.9 beam, about 600 mJy \citep{Gallimore96a},
 can only account for a small
fraction of the observed mid-IR emission \citep[see also][]{VM}.

\subsubsection{The nuclear silicate absorption feature}
\label{nucsil}

In the Michelle spectra of the central 0.4\arcsec\ core (Fig.~\ref{fig:specs}),
$\tau_{9.7}=0.39\pm 0.02$ for epoch 1 and $\tau_{9.7}=0.42\pm 0.02$ for
epoch 2. The N-band spectrum of NGC~1068 has also been measured by
\citet{Jaffe} in a 0.4\arcsec\ slit and by \citet{Rhee05} in a 0.25\arcsec\ $\times$ 0.25\arcsec\ aperture (Fig.~\ref{fig:specs}). We estimate $\tau_{9.7}\approx 0.30$
for the total flux spectrum of \citet{Jaffe}, in rough agreement with the results
presented here, although their spectrum has a somewhat shallower decline
towards shorter wavelengths than ours.  The \citet{Rhee05}  data bear a close resemblance to the Michelle spectra.

\begin{figure}[t]
\includegraphics[width=16cm,angle=0]{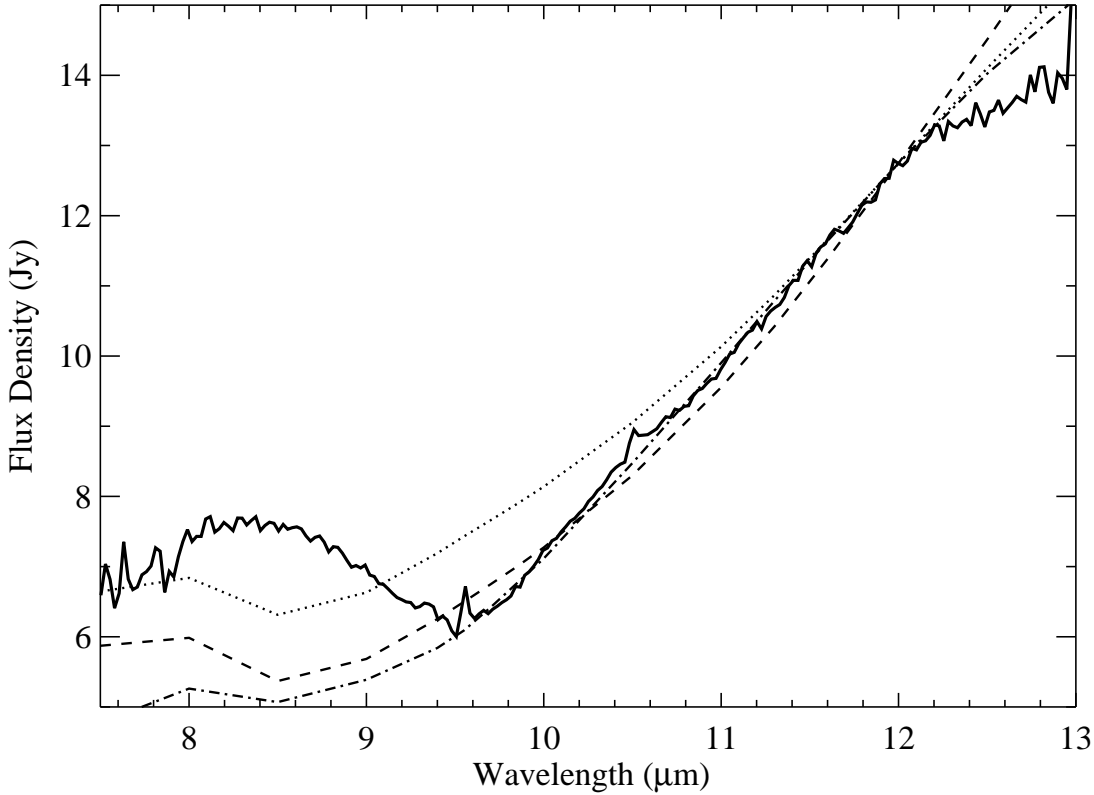}
\caption{Fits of the clumpy torus model to the 10 $\mu$m spectrum of NGC~1068
({\it heavy solid line}). Model spectra are scaled to the observed 12~$\mu$m
flux density and this scaling determines the AGN intrinsic luminosity, $L_{AGN} \approx
2\times 10^{45} {\rm \, erg\, s^{-1}}$. In all cases, $i=90^\circ$. The other
model parameters are: $N_c = 6$, $\tau_V=40$, and $q=2$ ({\it dotted line});
$N_c = 8$, $\tau_V=40$, and $q=2$ ({\it dashed line}); and $N_c = 20$,
$\tau_V=20$, and $q=3$ ({\it dot-dashed line}).
}
\label{fig:mod1}
\end{figure}

To gain insight into the dust distribution in the central 0.4\arcsec\
($\approx$ 30pc) of NGC~1068, we compare the nuclear spectrum with the
models of \citet{Nenkova}, who consider radiative transfer in clumpy
distributions of material.  In these models the angular distribution of
the clouds is a Gaussian centered on the equatorial plane. The clouds
follow a power law distribution in radius, $r$, with number declining as
$r^{-q}$. The emergent emission, calculated for all viewing angles, is a
function of ${N_c}$, the average number of clouds along an equatorial
radius, and $\tau_V$, the optical depth per cloud in the $V$ band.  In
fitting the models to the data, we characterize the continuum spectra in
terms of three quantities: $m_1$, the dimensionless slope from
8.0-9.0~$\mu$m, $m_2$, the dimensionless slope from 11.0-12.5~$\mu$m, and
$f_{8-12}$, the ratio of flux densities at 8 and 12 $\mu$m.  We considered a
coarse grid of $20 \le \tau_V \le 200$ and $5 \le N_c \le 20$. All the
model slopes tend to be steeper for smaller values of $q$. For a given
value of $q$, the slopes exhibit small ($10$\%) variations with viewing
angle. The variation in slope is more strongly correlated with $N_c$,
changing by $50$\% when $N_c$ changes by 5. Scaling a model to the
observed flux determines the intrinsic AGN luminosity, $L_{AGN}$.

In fitting the nuclear spectrum, we require inclination angle $i = 80$ or
$90^\circ$, as indicated by water masers \citep{Greenhill97}. With this
constraint, we find $\tau_{V} \le 60$ (Fig. \ref{fig:mod1}).  In nearly
all of the preferred models, $q = 2$ or 3, and $N_c \ge 5$.  The
best-fitting model parameters are $\tau_{V} = 40$, $q = 2$, and $N_c = 8$.
For $q=2$, a range of $N_c$ from 5 to 10 fits well, generally with
$\tau_{V} = 40$. The best $q=3$ models have $\tau_{V} = 20$, with $5 \le
N_c \le 20$. The cloud distribution is compact in these larger $q$ models.
Such a small region is consistent with high-resolution mid-IR images
\citep{Bock2,Tomo,Galliano05} and interferometry \citep{Jaffe} that show
almost no extended emission in the presumed torus mid-plane. For the
intrinsic AGN luminosity we find $L_{AGN} \approx 2\times 10^{45} {\rm \,
erg\, s^{-1}}$.

We specifically avoided fitting the silicate feature itself, and the resulting models do not reproduce its shape well. The resulting discrepancy is likely due at least in part to limitations of the current models. Also, the theoretical spectra account only for the torus's emission, but foreground absorption and contamination from extended sources within the aperture may alter the observed spectrum.

\begin{figure}
\includegraphics[width=16cm,angle=0]{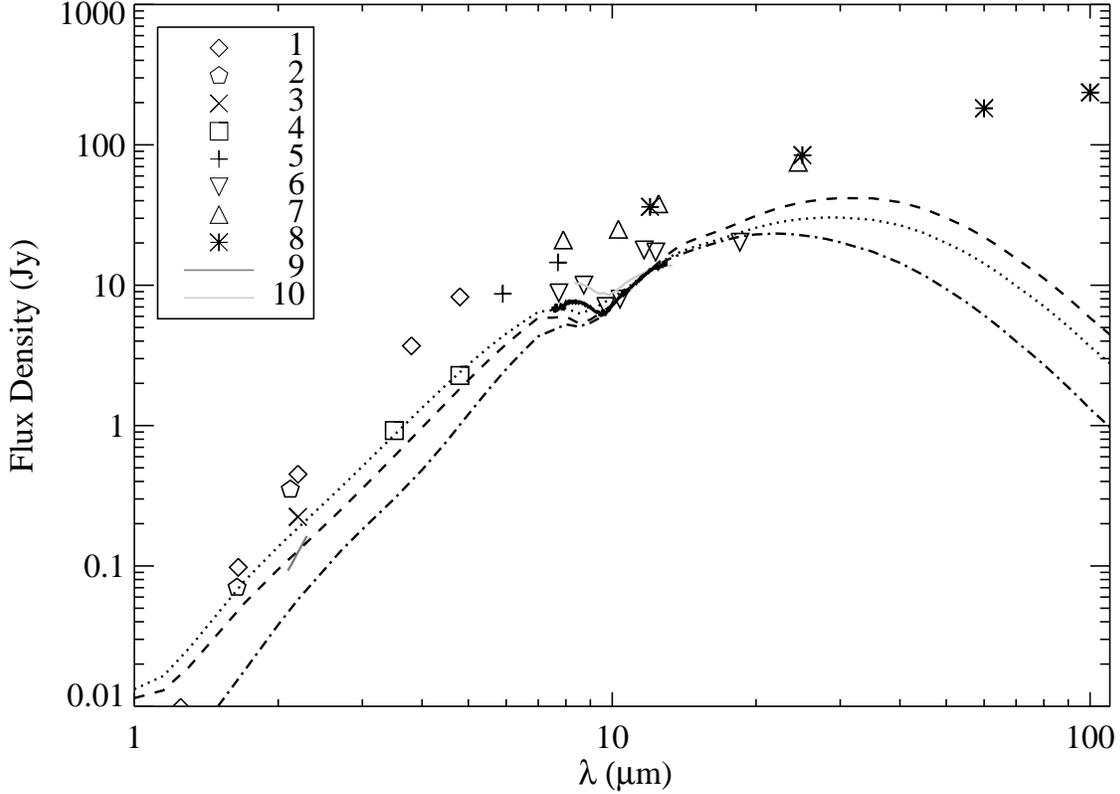}
\caption{Original data and models (as in Fig.~\ref{fig:mod1}) compared with
data from the literature.  The active nucleus alone contributes only
a fraction of the total luminosity of NGC~1068 that is measured on larger spatial scales.
Notes and references for the additional data:
{\bf 1}. Spatially unresolved within $1.5\arcsec$ radius;
\citet{Alonso01}.
{\bf 2}. $0.02\arcsec \times 0.04\arcsec $ compact core; \citet{Weigelt04}.
{\bf 3}. Spatially unresolved within $0.5\arcsec$ radius; \citet{Thatte97}.
{\bf 4}. $0.3\arcsec$ radius aperture; \citet{Marco00}.
{\bf 5}. Infrared Space Observatory,  $24\arcsec\times 24\arcsec$ aperture; \citet{Rigop}.
{\bf 6}. $0.29 \arcsec \times 0.18\arcsec$ aperture on deconvolved images; \citet{Tomo}.
{\bf 7}.  $2\arcsec$ radius aperture; \citet{Bock2}.
{\bf 8}. IRAS Survey,  typical aperture $1\arcmin \times 5\arcmin$; \citet{Soifer89}.
{\bf 9}. $0.3\arcsec$-wide slit; \citet{Gratadour03}.
{\bf 10}. $0.4\arcsec$-wide slit, integrated over the central emission peak only; \citet{Jaffe}.
}
\label{fig:mod2}
\end{figure}

Fig.~\ref{fig:mod2} shows the models from Fig.~\ref{fig:mod1} along with
the broader SED of NGC~1068 compiled from published results. The models
are plotted with the same scaling as in Figure 4 .
 Most of the measurements do not isolate the AGN and instead include a significant
contribution from the larger galaxy. Systematically lower fluxes, which
are more similar to the models, are observed only at the highest spatial
resolution. As already shown in Table 1, much of the observed mid-infrared
flux of NGC~1068 is not due to the active nucleus alone, even on
sub-arcsecond scales.

\subsubsection{The silicate feature in the ionization cones}
\label{NLR}

\begin{figure}[t]
\includegraphics[width=10cm,angle=-90]{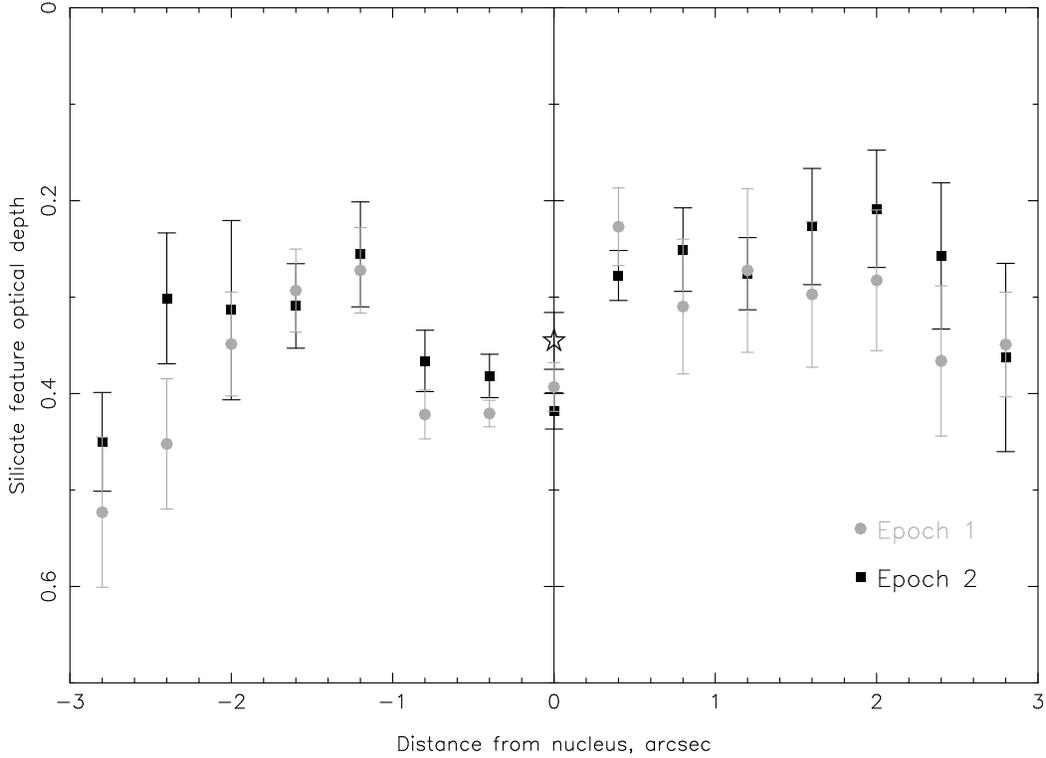}
\caption{$\tau_{9.7}$ as a function of distance from the nucleus in
0.4\arcsec\
bins. Positive distances are NNE; negative, SSW. The star indicates the
value
obtained from the total spectrum extracted over $\pm$3\arcsec\ along the
0.4\arcsec-wide slit.} \label{fig:tau}
\end{figure}

The complexity of the MIR emission from the ionization cones is
immediately apparent from Figures~\ref{fig:all} and \ref{fig:tau}, in
which neighboring spectra often exhibit significant variations in spectral
slope and silicate feature profile and depth, at positions both adjacent
to and more distant from the center. Smaller differences between the
spectra taken at the two epochs can also be seen. We attribute these to the
slight differences in seeing and slit positioning between the two data
sets, which therefore do not sample identical regions of the complex and
structured extended MIR emission.

Probably the most striking finding is the strong asymmetry in the spatial
distribution of the silicate "optical depth" close to the nucleus (Fig.~\ref{fig:tau}). The
largest values of $\tau_{9.7}$ occur 0--1\arcsec\ south-southwest (SSW) of the central
compact source. This finding is consistent with observations of optical
and near-IR line and continuum emission
\citep[e.g.][]{Macchetto94,Bruhweiler01,Thompson01}, both of which are
much fainter to the south of the nucleus, suggesting increased extinction
to this region. Further support for this interpretation comes from the
increase in near-IR polarized flux to the SSW, interpreted as additional
scattering in that area \citep{Young96,Packham97,Simpson02}. On the whole,
this picture is consistent with the dusty disk of the host galaxy, which is
inclined at $\sim$29\degr \citep{GG02}, screening the southern but not the
northern ionization cone, as suggested by a number of authors
\citep[e.g.][]{Packham97,Bock98,Kraemer00,Young01}.

We emphasize, however, that this conclusion could not have been drawn
from $\tau_{9.7}$ alone.  The complexity revealed by the observations, and
the likely clumpiness of the dust (\S\ref{sil}), make it difficult to
extract physical parameters in a meaningful way. The common interpretation
in terms of an absorbing screen in front of a background source ignores
the screen emission, becoming meaningless when optical depths are large.
The temperature of an externally heated optically thick dust cloud varies
considerably over its surface, and the depth of the absorption feature
emerging from its dark side is not a reliable indicator of optical depth
(Nenkova, Sirocky, Ivezi\'c \& Elitzur 2005, in preparation). The spectra
shown in figures \ref{fig:mod1} and \ref{fig:mod2} were calculated with
models in which the clouds' overall optical depth varied from 160 to 300
at V, yet they produce a feature with $\tau_{9.7}$ of only $\sim 0.3$. The
patchy structure found in both the shape and depth of the silicate feature
can be produced by many different combinations of clouds and background
continuum. The current observations do not provide the means for a unique
interpretation.


The profile of the silicate feature off the nucleus, especially to the north, deserves some comment. The spectra display clear signs of absorption but the shape of the feature is quite distinct from that of the typical diffuse ISM absorption band, which  commences close to 8~$\mu$m and
extends to $\sim$12.5~$\mu$m \citep[e.g.][]{Kemper04,Chiar05}. In Fig.~\ref{fig:all}, the feature has almost reached the continuum level by $\sim$11~$\mu$m in many of the spectra. Spectral profiles similar to those observed from 1.2 - 2.0\arcsec\ NNE have also been found in spectra
of embedded YSOs, and interpreted as a combination of absorption and emission in the line of sight \citep[][see for instance IRAS~04295+2251 in their Fig. 1]{K-S}. This may suggest that
the spectral shapes observed in the ionization cones of NGC~1068 indicate components of both absorption and emission from those regions.

In the Gemini data, the overall value of $\tau_{9.7}$  (i.e., obtained from the
sum of all spectra within 3\arcsec) is 0.35$\pm$0.03. This is slightly lower than
the values in the nuclear spectra. The overall value is also somewhat less than
that in the spectrum  of \citet{Sieb04} using a 3\arcsec\ slit of unspecified
orientation; employing the same methodology as that used for our silicate
absorption depth estimates, we measure $\tau_{9.7}\approx 0.45$ from their Fig.~14. \citet{Aitken84} and \citet{Roche84} observed NGC~1068 with beam sizes of 4
- 6\arcsec\ and estimated $\tau_{9.7}\approx0.5$.  Thus there may be a moderate
difference in $\tau_{9.7}$ between these measurements and our summed spectrum.
This could perhaps be accounted for if stronger silicate absorption is found in regions
outside of the 0.4\arcsec\ slit; if, for example, silicate emission from the
ionization cones fills in the absorption feature in the narrow slit.

The spatially-resolved mid-IR SED within 1\arcsec\ of the
mid-IR flux peak was previously examined by \citet{Bock2} and \citet{Galliano} using
imaging in four and seven narrow-band filters respectively. While
\citet{Bock2} detected prominent silicate absorption in the core and at a
position 0.4\arcsec\ south, and little absorption to the north,
\citet{Galliano} found silicate absorption increasing with distance from
the nucleus to the northeast, with the feature in emission to the southwest.  Thus, the
behavior of the silicate feature has until now been unclear. The findings
of \citet{Bock2} are in fair agreement with Figures \ref{fig:all} and
\ref{fig:tau}.  Contrary to \citet{Galliano}, neither we nor  \citet{Bock2}
observe silicate emission at positions south of the nucleus, finding
instead that the region within 1\arcsec\ of the nucleus has stronger
absorption at the flux peak and to the south than to the north. Similar results have recently been reported by \citet{Rhee05}. While the
maximum $\tau_{9.7}=0.42\pm0.02$ in our epoch 2 data is reached on the
nucleus itself, a feature of comparable depth
is observed at 0.4 and 0.8\arcsec\ SSW of the center in the epoch 1 spectra. The
silicate feature  strength
is fairly uniform at
$\tau_{9.7}\approx0.2-0.3$ in the northern ionization cone,
and  at 1.2-2.0\arcsec\ SSW of the
flux peak, $\tau_{9.7}$ in both sets of data falls to similar values to
that observed 
to the NNE.

\subsubsection{The emission lines.}
\label{lines}

\begin{figure}
\includegraphics[scale=0.7]{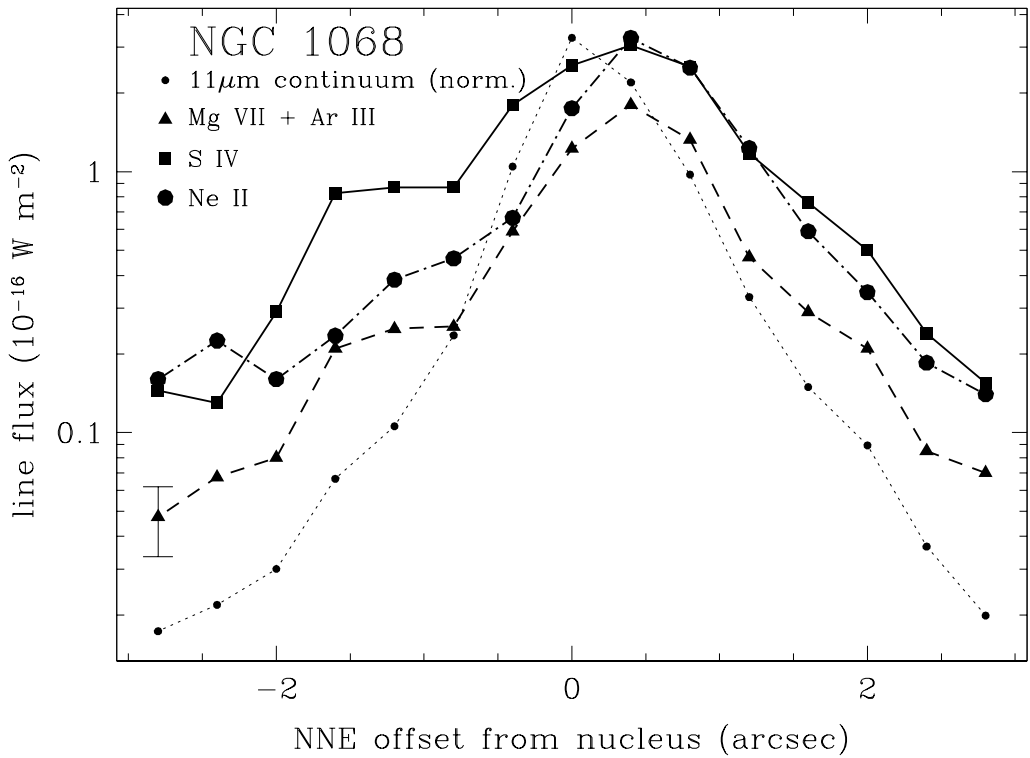}
\caption{Fluxes in a 0.4\arcsec\ x 0.4\arcsec\ aperture of three fine structure
lines  (mean of the two epochs) and the 11$\mu$m continuum flux density
as a function of offset from the nucleus.  A representative random error
bar ($\pm$1$\sigma$) is shown for the line fluxes; the uncertainty in the
continuum is negligible.}
\label{fig:lines}
\end{figure}

The [Ar~III]+[Mg~VII], [S IV], and [Ne II] line fluxes measured along the
slit are listed in Table \ref{table2} and plotted in Fig. \ref{fig:lines}.
As can be seen in the figure, although there are some differences between
them, all three are of fairly similar brightness at their peak (to within
a factor of two) and fall off rapidly to either side. The figure also
shows that the mid-infrared continuum is considerably more sharply
concentrated toward the nucleus than is the line emission.  None of the
fine structure line fluxes peaks at the nucleus.

Relative to the continuum peak there is a clear offset of ~0.2\arcsec\ in
the centroids of the fluxes of all of the fine structure lines relative
to the centroid of the continuum. This offset is to the north-northeast,
and is the direction expected based on the trends in extinction as a
function of position (see \S \ref{NLR}). In addition, all three line
fluxes peak 0.4\arcsec\ NNE of the center. Correcting the measured line
fluxes for extinction at each position along the slit based on
$\tau_{9.7}$ would result in more symmetric distributions about the
nucleus for all three lines. However, variations in extinction cannot
entirely explain the spatial distributions of the relative fluxes of the
three lines. For example, close to the center the large asymmetry in 
the [Ne~II] line flux at 0.4\arcsec\ NNE compared to 0.4\arcsec\ SSW will be
much less extinction-corrected than the asymmetries of the other lines,
whose wavelengths are closer to the peak of the silicate absorption. Thus other effects such as
variable excitation or an asymmetric distribution of material must also contribute to the offset between the continuum and emission line peaks. These may be related to narrow line clouds B and/or C \citep{Evans91}, located about 0.1-0.4\arcsec\ north of the nucleus and suggested to be responsible for the observed bend in the radio jet close to that position \citep{Gallimore96b,Galliano}.

\begin{deluxetable}{ccccc}
\tabletypesize{\scriptsize}
\tablecaption{Line fluxes and continuum flux density \tablenotemark{1,2}}
\tablewidth{0pt}
\tablecolumns{5}
\tablehead{
\colhead{NNE Offset} & \colhead{[Ar III + Mg VII] 9.0 $\mu$m} & \colhead {[S
IV] 10.5 $\mu$m} & \colhead{[Ne II] 12.8 $\mu$m} & \colhead{Continuum 11$\mu$m}\\
\colhead{(arcsec)} & \colhead{($10^{-16} \rm W m^{-2}$)} &\colhead{($10^{-16} \rm W m^{-2}$)} &\colhead{($10^{-16} \rm W m^{-2}$)} & \colhead{($ \rm W m^{-2} \mu m^{-1}$) }}
\startdata

 \sidehead{Epoch 1:}
2.8        &        0.09     &       0.17    &        0.14   & 1.59e-15\\
2.4        &        0.10      &      0.21     &       0.21  &  2.83e-15\\
2.0        &        0.21      &      0.45     &       0.29   & 6.47e-15\\
1.6        &        0.30      &      0.71     &       0.59   & 1.11e-14\\
1.2         &       0.53      &      1.15        &      1.31   & 2.73e-14\\
0.8        &        1.60       &      2.75        &      2.55  &  8.45e-14\\
0.4        &        2.35        &      3.50         &      3.10    & 1.85e-13\\
0.0        &        1.40        &      3.25        &       1.31   & 2.44e-13\\
-0.4       &         0.63      &      2.15     &        0.54  &  8.24e-14\\
-0.8       &         0.27      &      0.89      &      0.51 &  1.93e-14\\
-1.2       &         0.22      &      0.85      &      0.39   &  8.6 e-15\\
-1.6        &        0.22      &      0.96       &     0.27   &  5.5 e-15\\
-2.0       &         0.08      &      0.36       &     0.26 &   2.5 e-15\\
-2.4       &         0.06       &     0.15       &     0.25  &  1.80e-15\\
-2.8       &         0.05       &     0.18      &      0.17   & 1.4 e-15\\

\sidehead{Epoch 2:}
2.8         &       0.05    &       0.15      &      0.15  &  1.39e-15\\
2.4         &      0.07     &      0.27      &      0.16 &  2.65e-15\\
2.0         &       0.22      &      0.55       &     0.40  &  6.93e-15\\
1.6         &      0.29       &     0.81       &    0.59  &  1.13e-14\\
1.2          &     0.42     &       1.18     &       1.15  &  2.23e-14\\
0.8         &      1.05      &      2.30       &     2.45  &   6.16e-14\\
0.4         &      1.25       &      2.60        &     3.40   &  1.45e-13\\
0.0          &      1.05       &      1.85       &      2.15   &  2.44e-13\\
-0.4        &        0.56      &      1.45       &     0.80  & 7.43e-14\\
-0.8         &       0.24      &      0.86      &      0.42  &  1.61e-14\\
-1.2        &        0.28      &      0.89      &      0.39  &  7.25e-15\\
-1.6       &         0.20      &      0.69      &      0.20  &  4.48e-15\\
-2.0       &        0.08      &     0.23      &     0.17  &  2.00e-15\\
-2.4       &        0.08      &      0.12       &     0.19   & 1.48e-15\\
-2.8        &        0.05     &       0.12      &      0.15   & 1.20e-15\\

\enddata
\tablenotetext{1}{In a 0.4\arcsec x 0.4\arcsec aperture; line fluxes corrected for slit
loss in calibration star (see text)}
\tablenotetext{2}{ 2$\sigma$ uncertainties are 0.03 for [Ar III / Mg VII] and
        [S IV] and  0.04 for [Ne II]; flux calibration is accurate to $\pm$
        20\%.}
\label{table2}
\end{deluxetable}

Evidence for variable excitation is also present further from the
nucleus. The [Ne~II] and [S~IV] lines have roughly equal fluxes at their
peak, but at distances of 0.4-2\arcsec\ SSW of the nucleus the [S~IV] line
is typically 1.5-3 times the strength of each of the other two emission
features, which have similar spatial distributions. These differences are
too large to be attributed to the changes in extinction.

The extinction explanation is also untenable if the [Ar~III] line
dominates the flux of the 9.0~$\mu$m blend because the silicate
extinctions at 9.0~$\mu$m and 10.5~$\mu$m are similar. However, the
[Mg~VII] line is expected to contribute significantly to the strength of
the 9.0~$\mu$m feature near the nucleus, as lines of many highly excited
species are seen toward the nucleus, including a strong 5.50~$\mu$m line
of [Mg~VII] \citep{Lutz00}. The change in the ratio of the 9.0~$\mu$m and
10.5~$\mu$m features can be understood as a consequence of the rapid
fall-off of the coronal line flux with distance from the nuclear ionizing
source and the increased importance of other ionizing sources such as hot
stars. The wavelength of the [Ar~III] line (8.991~$\mu$m)is accurately
known and the best estimate for the [Mg~VII] line wavelength places it
within 0.004~$\mu$m of the [Ar~III] line \citep{Hayward96}. Thus it will
be difficult for higher spectral resolution to separate the contributions
of the two lines and indeed the R$\sim$1500 spectrum of \citet{Lutz00}
does not.

The strength of the 9.0~$\mu$m blend near the nucleus, roughly half that of
the other two lines, also suggests that [Mg~VII] (with an ionization
potential of 186.5~eV) is a significant and probably dominant contributor
to this feature. This inference is based on the observation that in starburst
galaxies the [Ar~III] line is always at least a factor of four weaker than
at least one of the other two lines \citep{Verma03}.  That is
understandable because (1) argon is 30 times less abundant than neon and
almost an order of magnitude less abundant than sulfur, (2) the fine
structure collision strengths for excitation of the [S~IV] and [Ar~III]
transitions' upper levels are comparable and only several times larger
than that for the [Ne~II] upper level, and (3) the ionization potential of
[Ar~III] (27.6~eV), is intermediate between those of [Ne~II] (21.6~eV) and
[S~IV] (34.8~eV).

Comparison of the integrated fluxes of the [Ar~III]+[Mg~VII], [S~IV], and
[Ne~II] lines along the central 6\arcsec\ of the 0.4\arcsec\ Michelle slit
with the fluxes measured in the 14\arcsec\ $\times$ 20\arcsec\ slit of the
ISO SWS \citep{Lutz00} demonstrates that the bulk of the emission line
flux detected by ISO SWS comes from the central 1-2 arcseconds. The fluxes
detected by Michelle are approximately 28\%, 27\%, and 18\%, respectively,
of those detected by ISO SWS. In their 3\arcsec\ slit, \citet{Sieb04}
observed 52\% of the ISO SWS [S IV] flux. Compared with the 28\% that we
detect, this suggests that the [S~IV] flux is quite diffuse within the
ionization cones, as is indeed illustrated by Fig. ~\ref{fig:lines}. Both our narrow-slit data and the wider-slit measurements of  \citet{Sieb04} recover only $\la$20\% of the large-aperture
[Ne~II] flux, consistent with the more centrally-concentrated nature of
the {\em nuclear} [Ne~II] emission shown in Fig.~\ref{fig:lines} and suggesting that the bulk of the [Ne II] flux  detected by ISO arises in more distant regions, such as the kpc-scale starburst ring, not covered by these slits.

\subsection{The M-band spectrum}

\begin{figure}[t]
\plotone{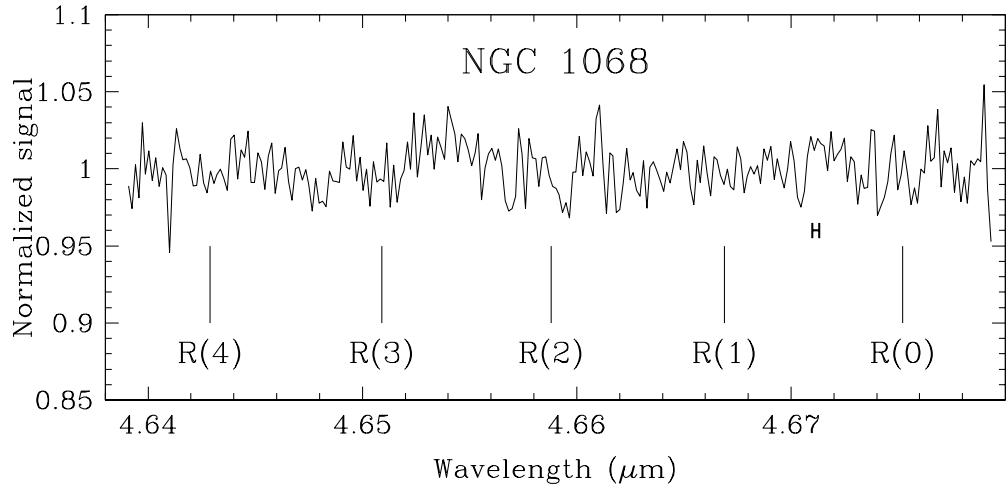}
\caption{Spectrum of NGC~1068  near 4.66~$\mu$m. The velocity resolution
of 20 km/s is indicated. The positions of CO lines are shown for
the redshift of NGC~1068, z=0.00379.}
\label{fig:CO}
\end{figure}

The UKIRT/CGS4 R=15000 M-band spectrum of NGC~1068 is shown in Fig.
\ref{fig:CO}. The spectrum covers a narrow portion of the M window that
contains the redshifted R(0) - R(4) lines of the fundamental
vibration-rotation band of $^{12}$CO.  In absorption the R(0) line arises
from the lowest lying rotational level, the lower state energies of the
other transitions correspond to temperatures of 6 K, 17 K, 33 K, and 55 K
above the ground state. In LTE at least some of these lines would be among
the strongest CO absorption lines for gas temperatures up to several
hundred Kelvins. In dense ($n$$>$10$^5$~cm$^{-3}$) clouds CO should be in
near-LTE at these temperatures. In lower density clouds, because
spontaneous emission is more rapid than collisional excitation, LTE will
not be achieved and most of the CO will be in the lowest one or two
rotational levels despite a high gas kinetic temperature \citep[see
e.g.][]{Geballe99, McCall02}.

No CO lines are apparent in Fig.  \ref{fig:CO}.  The limit per 20 km/s
resolution element is typically 3\% of the continuum. \citet{LutzCO} have
published an upper limit of 7\% of the continuum for individual lines, based on
an R=2500 spectrum obtained using ISO-SWS. The much higher resolution spectrum
presented here thus provides a considerably more stringent limit, unless the CO
lines are several hundred km/s wide. 

As discussed in \S\ref{nuc}, there are several ways of interpreting this upper
limit. Here we analyze it in terms of constraining the properties of a
foreground screen, although this is surely a naive interpretation. For
such an absorber the limit on the CO column density would depend on the CO excitation temperature and the linewidth.
For 50~K and 50~km/s, for example, the UKIRT spectrum implies that the
total CO column density is less than $\sim 6 \times 10 ^{15} {\rm
cm}^{-2}$. Higher temperatures and greater linewidths both increase that
upper limit (but only by a factor of about two for 100 K and 100 km/s).

If the CO-containing gas (as well as the associated obscuring dust) is in
dense clouds (i.e., clouds into which near-UV radiation cannot penetrate),
then almost all of the interstellar carbon is in CO. Assuming that the
carbon abundance is roughly solar, one concludes that N(H$_{2}) < 1 \times
10^{20}~\rm{cm}^{-2}$.  For a standard gas-to dust ratio, the corresponding
foreground extinction is $\sim$0.1 mag.
If the obscuring material is entirely in diffuse clouds, then at most only a few
percent of the carbon is neutral and the H$_{2}$ column density and dust
extinction can be roughly two orders of magnitude higher.

\section{Discussion}
\label{discuss}

\subsection{Clumpiness and complex spectral structure in the ionization cones}
\label{sil}

Perhaps the most striking aspect of these data is the variation in the spectra over small distances (Fig.~\ref{fig:all}). There
is significant structure in the dust and gas in the circumnuclear regions even on
subarcsecond scales: the ionization cones of NGC~1068 are as complex in
the mid-IR as at shorter wavelengths. This complexity was hinted at in the
narrow-band imaging study of \citet{Bock2} and is evident in the
deconvolved 12.8~$\mu$m image of \citet{Galliano05}, and is made
clearer in the present data.

As noted in \S\ref{NLR}, the bulk of the mid-IR emission coincident with
the ionization cones in NGC~1068 is likely to be thermal emission from
warm dust. 
Any arrangement of
the dust proposed to explain the mid-IR emission and silicate feature,
however, must also be consistent with certain tight observational
constraints. First,
the reddening of the northern ionization cone is lower than suggested by
the depth of the silicate feature in our spectra. If a foreground screen
distribution and standard Galactic dust properties
\citep[$A_{V}/\tau_{9.7}\approx9-19$ and $R_{V}\approx 3$;
][]{RA84,Roche85,Whittet92} were assumed, 
 $\tau_{9.7}=0.2$ would imply $E_{B-V}\sim0.6-1.3$, already somewhat higher than the mean reddening of the northern cone, $E_{B-V}\approx 0.35$ \citep[for the
more obscured blueshifted line components;][]{Kraemer00}. Moreover, as the
spectral profile of the silicate feature in the cones may reflect a
combination of both emission and absorption, any determination of $E_{B-V}$
from $\tau_{9.7}$ in this simple screen model could easily be an
underestimate. Second, the UV/optical polarisation mechanism in the
central $\la$1\arcsec\ of NGC~1068 is thought to be electron scattering
\citep{AM85,MGM,AHM94,Young95}, mainly on the basis of the
wavelength-independent polarization spectrum, but the high efficiency for
dust relative to electron scattering \citep[e.g.][]{Kishimoto01} suggests that light scattered off
electrons in the cones should undergo further, wavelength-dependent
scattering off any dust grains in or around the cones.

Arranging the dust in optically thick clumps associated with the NLR
clouds, in accordance with the good spatial coincidence of the structures
emitting in [O III] and at 12.8~$\mu$m \citep{Galliano05}, may circumvent
these problems. Gray scattering from such clumps may be able to mimic the
wavelength-independent polarization caused by electron scattering
\citep{Code95,Kishimoto01}, and the dust would be likely to survive for
much longer in this kind of configuration \citep{Laor93,VM}. Arranging the
dust in clumps could also explain the low reddening of the ionization
cones while permitting appreciable silicate absorption, as much of the
ionized gas would be expected to lie outside the clumps, which would only
have a small covering factor.  In this scenario, any
emission component in the silicate feature in the cones would result from
a view of the hot sides and faces of clumps towards the far edge of the
cone. The absorption would then arise in the cooler surfaces of clumps
closer to the near edge of the cone, inclined away from the central
engine, and in a small amount of foreground dust. The presence of dust clumps was suggested by
\citet{Galliano} to explain SEDs showing silicate emission in the southern ionization cone (stated to be tilted away from the line of sight and therefore exposing hot clump surfaces to our view), at odds with the present data and \citet[][]{Bock2}. However, the southern cone is also thought to be located behind the disk of the galaxy and subject to increased extinction that may well suppress a silicate emission feature from hot clump faces. Further from the nucleus, on the other hand  --- around and including the "northeast knot" first
identified by \citet{MGM} --- both dust and electron scattering are
thought to contribute to the polarization \citep{MGM,Packham97,Simpson02},
suggesting the presence of dust other than in clumps. 


\subsection{The nature and distribution of material close to the nucleus}
\label{nuc}

In fitting both the nuclear 10~$\mu$m spectrum and the broader IR SED, the
clumpy torus models require most of the mid-IR-emitting dust in the torus
to exist rather close to the nucleus. In the models with $q=3$, for
example, 90\% of the clouds lie within 3 pc of the AGN. This is comparable
to the dimensions of the warm dust structure detected by \citet{Jaffe} in
NGC~1068, and also in agreement with recent imaging data constraining the
central IR source to a diameter of a few parsecs or less in several other
Seyfert galaxies \citep{Swain03,Prieto04,Prieto05,Packham05}.

On the other hand, both mid-IR imaging and the torus model predictions imply
that substantial additional mid-IR emission arises outside the central source.
This is also apparent in the comparison of observational data in different
apertures shown in Fig.~\ref{fig:mod2}: there is a clear trend for
subarcsecond-aperture measurements at wavelengths as short as 3~$\mu$m to be
appreciably lower than larger-aperture ones, even when stellar dilution is
taken into account in the 3-5~$\mu$m data, as in \citet{Alonso01}. It is
important to note the wavelength dependence of trends among different
apertures. In the near-IR, the 1.5\arcsec\ measurements of \citet{Alonso01} are
in good agreement with the high-resolution results of \citet{Weigelt04} at both
H and K bands. The compact emission is known to dominate at these wavelengths,
contributing about 80\% of the flux in a $2.5 \times 2.5\arcsec$ aperture at K
\citep{Weigelt04}. In the L and M bands, however, the \citet{Alonso01} flux density
measurements exceed the compact source by a large factor, and this difference
persists in the \citet{Bock2} 2\arcsec-aperture N band measurements. While
estimates of the torus emission are known to be unreliable in the far-IR
because of the large beams involved, even ground-based mid-IR observations do not usually
isolate the AGN contribution.

This is also relevant to the interpretation of the non-detection of CO lines in
the M band spectrum: the M-band imaging of \citet{Rouan04}, for example,
implies that the aperture used for the CO observations contains emission from
much material not directly associated with the torus.  We therefore first
examine the limiting case in which all of the gas exists in optically thick
clouds much smaller than the beam size --- such as the torus clouds modeled in
\S\ref{nucsil} and the NLR clouds proposed in \S\ref{sil} --- and then go on to
consider the constraints the CO non-detection places on the nature and
distribution of material both outside and inside those clumps.

If the only molecular material is in optically thick molecular
clouds, the upper limit on the CO lines can be interpreted as constraining
the filling factor, $\phi$, of these clouds. The dark sides of individual
clouds will be strong CO absorbers, so we assume that the absorption from such
clouds removes all the light from the beam at that position. The limit on CO
line absorption then implies that the beam filling factor of molecular clouds
must be less than 0.03, comparable to the $\phi\sim$0.1 quoted by \citet{Nenkova}
as a reasonable realization of the clumpy torus model. On the other hand, if the entire
5~$\mu$m continuum source including the torus clouds is obscured in part by
cool foreground material, the implications of the non-detection of CO will
depend on the nature of that material. As discussed earlier, in dense molecular
clouds essentially all of the C is in CO, and the limit on the CO column
density in implies much less than a magnitude of visual foreground extinction.
In diffuse clouds typically only 1\%\ of the C is in CO, so that the limit can
tolerate as much as 5-10 mags of visual foreground extinction.

As noted in \S\ref{NLR}, in realistic clumpy dust configurations, however,
the relation between $\tau_{9.7}$ and optical depth is not
straightforward. Therefore Fig.~\ref{fig:tau} may not be appropriate for
estimating the extinction to the 5~$\mu$m continuum emitting regions. The
$E_{B-V}$ of the region covered by the 0.9\arcsec\ M band slit has been
derived from He II line ratios by \citet{Kraemer00}, who find a mean value
of 0.22 to the NNE and values typically $>$0.3 to the SSW. Using the
standard value of $R_{V}$, this translates to $A_{V}\sim$0.7-0.9 or more.
This is several times greater than the extinction implied by the CO
non-detection assuming the foreground material to be molecular in nature,
but consistent with the presence of diffuse clouds.

There is further evidence that supports diffuse clouds as the dominant
location of the dust: the presence of the 3.4~$\mu$m absorption feature in
the spectrum of the nucleus of NGC~1068
\citep{Wright96,Imanishi97,MB03,Me2}. The 3.4~$\mu$m absorption feature is
believed to be produced by hydrocarbons in refractory dust grains and has not been detected in dense clouds in the Galaxy, but has been
seen in numerous diffuse Galactic clouds. The absence of the feature in dense
clouds is thought to be at least partly due to the lack of atomic hydrogen
available to re-form C--H bonds destroyed by UV photons and cosmic rays
\citep{Munoz01,Mennella03}.

In the Galaxy the ratio of the total extinction to the optical depth of
the 3.4~$\mu$m feature has been measured in a large number of sources
\citep{P94,Rawlings03}. The value of $A_V$/$\tau_{3.4}$ is $\ga$250 in
diffuse clouds outside of the Galactic center and $\sim$150 toward the
Galactic center (where it is believed that roughly one-third of the
extinction arises in dense cloud material, making the ratio $\sim$100 for
diffuse clouds along that sightline).  In NGC~1068, the optical depth of
the 3.4~$\mu$m feature is measured to be about 0.1-0.15
\citep{Wright96,Imanishi97,MB03,Me2}. If the visual extinction derived
from the He~II lines is applicable to the regions producing the 5~$\mu$m
continuum the ratio is $<$10, and even if one derives an extinction from
$\tau_{9.7}$ the ratio is $\sim$50. Thus, the 3.4~$\mu$m feature is far
stronger relative to the foreground visual extinction in NGC~1068 than in
the Galaxy. These low values suggest that not only the foreground
extinction but much of the cloud material, regardless of its location,
giving rise to the silicate feature in the central 1\arcsec\ is diffuse in
terms of its chemistry and abundances.

\section{Conclusions}
\label{conc}

N-band near-diffraction-limited observations of the nuclear region of
NGC~1068 at the Gemini North telescope show that the complexity of the
nuclear region, previously seen in mid-IR imaging and observations at
shorter wavelengths, persists in mid-IR spectra. The spectra vary
widely on spatial scales of less than 1\arcsec, in the slope of the
continuum, the strength of the silicate absorption feature, and the fine
structure line fluxes. The optical depth of the silicate feature is
observed to vary between about 0.2 and 0.45,
and the silicate absorption and all three of the [Ar~III]+[Mg~VII], [S~IV]
and [Ne~II] fine structure features are asymmetrically distributed within
1\arcsec\ of the center. The relative strengths of the above three
emission features imply a strong contribution by the high excitation [Mg~VII]
line to the blend of that line and [Ar~III].

A high spectral resolution UKIRT/CGS4 spectrum of the central 1\arcsec\ of NGC~1068
shows no evidence for absorption by CO. This upper limit implies either a
small filling factor for dense molecular clouds obscuring the nuclear
continuum source(s), a predominance of obscuring diffuse
cloud material lacking in CO (whose existence also is implied by the interstellar
3.4~$\mu$m absorption feature seen towards the nucleus), or a combination
of the two.

The mid-IR emission from NGC 1068 originates in two distinct components.
Emission from the central region with diameter 0.4\arcsec\ is dominated by the
IR - bright obscuring torus. Our observations place an upper limit of 15 pc on
the torus radius, in agreement with the compact ($\sim$3 pc) size determined by
mid-IR interferometry. We have been able to obtain good fits with the clumpy
torus models of \citet{Nenkova} to this nuclear region. Mid-IR emission in
apertures larger than $\sim$ 1\arcsec\ is dominated by a second component, the
lower brightness, diffuse emission from the AGN-heated dust in the ionization
cones. If NGC 1068 is representative of other sources, the implication is that
AGN SEDs are highly uncertain since the torus does not dominate large-aperture
photometry at wavelengths longer than the near IR. While this has long been
known to be the case for facilities such as ISO and IRAS, our results underline
the need for future thermal-IR observations of AGN to be obtained at the
highest possible spatial resolution even at mid-IR wavelengths.

\acknowledgments
We wish to thank W. Jaffe and J. Rhee for providing data, and the anonymous referee for some sensible suggestions. CP acknowledges work supported by the National Science Foundation under Grant    
No. 0206617. NAL acknowledges work supported by NSF grant 0237291.
This paper is based in large part on observations obtained at the Gemini Observatory,
which is operated by the Association of Universities for Research in
Astronomy, Inc., under a cooperative agreement with the NSF on behalf of
the Gemini partnership: the National Science Foundation (United States),
the Particle Physics and Astronomy Research Council (United Kingdom), the
National Research Council (Canada), CONICYT (Chile), the Australian
Research Council (Australia), CNPq (Brazil) and CONICET (Argentina). The
United Kingdom Infrared Telescope is operated by the Joint Astronomy
Centre on behalf of the U.K. Particle Physics and Astronomy Research
Council.



\end{document}